\documentclass[aps,nofootinbib,showpacs,showkeys,preprint
] {revtex4}

\usepackage{epsf,epsfig,subfigure,axodraw,graphicx,amsmath,amssymb}
\usepackage{color}

\def\lsim{\lower.7ex\hbox{$\;\stackrel{\textstyle<}{\sim}\;$}}

\def\Qem{{$Q_{\rm em}$}}

 \def\Z{{\bf Z}}

\def\EE{E$_8\times$E$_8^\prime$}

 \def\Z{{\bf Z}}

\def\EE{E$_8\times$E$_8^\prime$}
\def\Eo{E$_8$}
\def\Eh{E$_8'$}

\def\two{{\bf 2}}
\def\fiveb{\overline{\bf 5}}
\def\five{{\bf 5}}

\def\ten{{\bf 10}}
\def\one{{\bf 1}}

\def\threeb{\overline{\bf 3}}
\def\three{{\bf 3}}

\begin{document}
\preprint{SNUTP 07-010}
\title{GMSB at a stable vacuum and MSSM  without exotics from
heterotic string}
\author{Jihn E.  Kim
}
\address{Department of Physics and Astronomy and Center for Theoretical
 Physics, Seoul National University, Seoul 151-747, Korea }

\begin{abstract}
 {We show that it is possible to introduce the confining hidden sector
gauge group SU(5)$'$ with the chiral matter  $\ten'_0$ plus
$\fiveb'_0$, which are neutral under the standard model gauge group,
toward a gauge mediated supersymmetry breaking (GMSB) in a
$\Z_{12-I}$ orbifold compactification of \EE\ heterotic string.
Three families of MSSM result without exotics.  We also find a
desirable matter parity $P$ (or $R$-parity) assignment. We note that
this model contains the spectrum of the Lee-Weinberg model which has
a nice solution of the $\mu$ problem.
  }
\end{abstract}


 \maketitle

\section{Introduction}

The supersymmetric (SUSY) extension of the standard model (SM)
encounters a few naturalness problems, the SUSY flavor problem
\cite{Masiero}, the little hierarchy problem \cite{Little}, the
$\mu$ problem \cite{Mu}, etc. The hierarchichal magnitude is worst
in the $\mu$ problem but here there are nice solutions \cite{MuSol}.
The little hierarchy problem has weakened the nice feature of the
SUSY solution of the gauge hierarchy problem and we hope that it
will be understood somehow in the future. On the other hand, the
SUSY flavor problem seems to require family independence of the
interactions at the GUT scale. The attractive gravity mediation
scenario for transmitting SUSY breaking down to the observable
sector probably violate the flavor independence of interactions
violently. This observation has led to the gauge mediated
supersymmetry breaking (GMSB) \cite{DineNeslson}. However, the
superstring attempt toward a GMSB model has not been successful
phenomenologically, even though the possibility of SUSY breaking
spectra was pointed out \cite{Kim07}.

Recently, dynamical SUSY breaking (DSB) at an unstable minimum at
the origin of the field space got quite an interest following
Intrilligator, Seiberg and Shih (ISS)
\cite{ISS,Murayama06,ISSfollow}, partly because it has not been
successful in deriving a phenomenologically attractive model in the
stable vacuum. Among the results on SU($N$), SO($N$) and Sp($2n$)
groups, the result is especially simple for SU($N_c$) with $N_f$
flavors, showing an unstable minimum for $N_c+1\le N_f<\frac32 N_c$.
This mechanism is easily applicable to SU(5)$'$ models with 6 or 7
flavors, which can be realized in string compactifications
\cite{Kim07}.  Nevertheless, it is better to realize a
phenomenologically successful SUSY breaking {\it stable minimum},
not to worry about our stability in a remote future. In this paper,
therefore, we look for a GMSB spectrum in the orbifold
compactification of the \EE\ heterotic string with three families,
trying to satisfy all obvious phenomenological requirements.

The well-known DSB models are an SO(10)$'$ model with $\bf 16'$ or
$\bf 16'+10'$ \cite{SO10}, and an SU(5)$'$ model with $\bf
10'+\overline{5}'$ \cite{ADS}. It is known that GMSB with $\bf
16'+10'$ can be obtained from heterotic string \cite{KimKyaegut},
but the beta function magnitude is too large (in the negative) so
that SO(10)$'$ confines somewhat above $10^{13}$ GeV against a
meaningful GMSB. If the hidden sector gauge group is large, the
content of matter representation is usually small and the beta
function magnitude (in the negative) turns out to be too large to
implement the GMSB scenario. If the confining group is SU(4)$'$ or
smaller, it is not known that one can obtain a SUSY breaking stable
minimum. Thus, SU(5)$'$ is an attractive choice for the GMSB
\cite{Kim07}. To solve the SUSY flavor problem along this line of
the GMSB, we require two conditions: {\it relatively low hidden
sector confining scale} ($\lesssim 10^{12}$ GeV) and  {\it
appearance of matter spectrum allowing SUSY breaking}.

A nice feature of the ISS type model at an unstable vacuum toward
model building is that the SUSY breaking can be mediated through
dimension-4 superpotential given in\footnote{This form has been
considered by many \cite{ISSfollow}, in particular in
\cite{Murayama06}.}
 $$
 W\sim \frac{1}{M}Q\overline{Q} f\bar{f}
 $$
where $Q$ is a hidden sector quark and $f$ is a messenger. It is
possible because the vectorlike representations, for example six or
seven $(Q+\overline{Q})$, are present and the $Q\overline{Q}f \bar
f$ interaction is suppressed by one power of mass parameter. So this
mass parameter can be raised up to the GUT scale.

On the other hand, the uncalculable model with $\ten'+\fiveb'$ of SU(5)$'$
does not have such a simple singlet direction in terms of chiral
fields. For example, the term
$\epsilon_{ijklm}10^{ij}10^{kl}10^{mn}\bar 5_n=0$ since taking
$n=1$ without generality it is proportional to
$\epsilon_{1jklm}10^{1j}10^{kl}10^{m1}\bar 5_1$ which can be shown to
be vanishing using the antisymmetric symbol $\epsilon$. The singlet
combination is possible in terms of the chiral gauge field strength,
${\cal W'}^\alpha{\cal W'}_\alpha$.
 It is pointed out that the $F$-term of this singlet
combination can trigger the SUSY breaking to
low energy \cite{Murayama07},
 $$
 {\cal L}=\int d^2\theta \left( \frac{1}{M^2}f\bar{f}
 {\cal W'}^\alpha{\cal W'}_\alpha+M_f f\overline{f}\right)+{\rm h.c.}
 $$
 where the effective parameters of $M$ and $M_f$ can be lower than the
 GUT scale.

The GMSB problem in string models is very interesting. For example,
quite recently but before ISS, it has been reviewed \cite{GMSBst},
but the phenomenological requirements toward the minimal
supersymmetric standard model (MSSM) have made it difficult to be
found in string models. The three family condition works as a strong
constraint in the search of the hidden sector representations. If we
require the exotics free condition, the possibility reduces
dramatically.

In a $\Z_{12-I}$ orbifold compactification, we find a model
achieving the GMSB at a stable vacuum together with three families
of quarks and leptons without any exotics. Since there is no
exotics, it is hoped that the singlet VEVs toward successful Yukawa
couplings have much more freedom, most of which are set at the
string scale. We find a successful embedding of matter parity $P$
and a nice solution of the $\mu$ problem. One unsatisfactory feature
is that $\sin^2\theta_W$ is not $\frac38$. Thus, to fit the weak
mixing angle to the observed value, we must assume intermediate
state vectorlike particles. Anyway, another kind of intermediate
state particles is needed also for a successful messenger mass
scale.

\section{A $\Z_{12-I}$ Model}\label{sec:Model}

The twist vector in the six dimensional (6d) internal space is
\begin{align}
{\rm \Z_{12-I}\ shift}:\quad
\phi&=\textstyle(\frac{5}{12}~\frac{4}{12}~\frac{1}{12}).
\end{align}
We obtain the 4D gauge group by considering massless conditions
satisfying $P\cdot V=0$ and $P\cdot a_3=0$ in the untwisted sector
\cite{DHVW:1985}. We embed the discrete action $\Z_{12-I}$ in the
\EE\ space in terms of the shift vector $V$ and the Wilson line
$a_3$ as
\begin{align}
 V&=\textstyle
 \frac1{12}(6~6~6~2~2~2~3~3)(3~3~3~3~3~1~1~1)'
 \label{Z12ImodelC}\\
  a_3&=\textstyle
\frac{1}{3}(1~1~ 2~0~0~0~0~0)(
 0~0~0~0~0~1~1~{\textstyle -2})'.\label{WilsonLine}
\end{align}

\vskip 0.5cm \noindent {\bf  (a) Gauge group}:
 The 4D gauge groups are obtained by $P^2=2$ vectors
satisfying $P\cdot V=0$ and $P\cdot a_3=0$ mod integer,
\begin{align}
SU(3)_c\times &SU(3)_W \times SU(2)_n\times U(1)_a \times
U(1)_b\times U(1)_c\nonumber \\
&\times [SU(5)'\times SU(3)'\times U(1)^{'\ 2}].\label{gauge}
\end{align}
The gauge group SU(3)$_W$ will be broken down to SU(2)$_W$ by the
vacuum expectation value (VEV) of $\three$ and $\threeb$ of
SU(3)$_W$. Then, the simple roots of our interest SU(3)$_c$,
SU(2)$_W$, and SU(2)$_n$ are
\begin{align}
&{SU(3)_c:}\left\{
\begin{array}{cc}
\alpha_1=& (1\ {\textstyle -1}\ 0\ 0\ 0\ 0\ 0\ 0)\\
\alpha_2=& (0\ 1\ 1\ 0\ 0\ 0\ 0\ 0)
\end{array}\right. \label{SU3croots}\\
&{SU(2)_W:}\left\{
\begin{array}{cc}
\alpha_1=& ( 0\ 0\ 0\ 1\ {\textstyle -1}\ 0\ 0\ 0)
\end{array}\right. \label{SU3Wroots}\\
&{SU(2)_n:}
\begin{array}{cc}
\alpha_1=& ( 0\ 0\ 0\ 0\ 0\ 0\ 1\ {\textstyle -1})\\
\end{array}.
\end{align}
The hypercharge direction is the combination of U(1)s of Eq.
(\ref{gauge}) and some generators of nonabelian groups
\begin{equation}
Y=Y_{\rm Abel}+\frac1{\sqrt3}W_8+F_3-\frac1{\sqrt3}F_8=\tilde
Y+F_3-\frac1{\sqrt3}F_8\label{Ydef}
\end{equation}
where
\begin{equation}
Y_{\rm Abel}=Y_8+Y_8',\label{Yabel}
\end{equation}
and $W_8, F_3, F_8$ are nonabelian generators of SU(3)$_W$ and
SU(3)$'$. We define $\tilde Y=Y_{\rm Abel}+\frac1{\sqrt3}W_8$ by
including the U(1) generators of SU(3)$_W$ and SU(2)$_V$ (by VEVs of
scalar fields). $Y_8$ and $Y_8'$ are a linear combination of three
U(1) generators in \Eo\ and a linear combination of two U(1)
generators in \Eh, respectively. $W_8$ is the eighth generator of
SU(3)$_W$, $(\frac1{2\sqrt3}, \frac1{2\sqrt3}, -\frac1{\sqrt3})$,
and $F_3$ and $F_8$ are the third and the eighth generators of
SU(3)$'$, $(\frac12,-\frac12,0)$ and $(\frac1{2\sqrt3},
\frac1{2\sqrt3}, -\frac1{\sqrt3})$, respectively. We find that
exotics cannot be made vectorlike if we do not include $Y'$. $\tilde
Y$ is defined as
\begin{equation}
\tilde Y=Y_{\rm Abel}+\frac1{\sqrt3}W_8=\textstyle
(\frac{1}{6}~\frac{1}{6}~\frac{-1}{6}~0~0~\frac{-1}{2}~
\frac{-1}2~\frac{-1}2)(\frac12~\frac12~\frac12~
\frac12~\frac12~\frac{-1}{6}~\frac{-1}{6}~\frac{-1}{6})'.
\label{tildeY}
\end{equation}
We included the SU(3)$'$ generators in $Y$ of (\ref{Ydef}) so that
there does not appear exotics.

The five U(1) generators of (\ref{gauge}) are defined as
\begin{align}
\begin{array}{l}
Q_1= (6~6~\textstyle{-6}~0~0~0~0~0)(0~0~0~0~0~0~0~0)' \\
Q_2=  (0~0~0~6~6~6~0~0)(0~0~0~0~0~0~0~0)' \\
Q_3=  (0~0~0~0~0~0~2~2)(0~0~0~0~0~0~0~0)'\\
Q_4=  (0~0~0~0~0~0~0~0)(4~4~4~4~4~0~0~0)'\\
Q_5= (0~0~0~0~0~0~0~0)(0~0~0~0~0~4~4~4)'
\end{array}\label{U1s}
\end{align}

\vskip 0.5cm \noindent {\bf  (b) Matter representations}:
 Now there is a standard
method to obtain the massless spectrum in $\Z_{12-I}$ orbifold
models. The spectra in the untwisted sectors $U_1, U_2,$ and $U_3$,
and twisted sectors, $T1_{0,+,-},T2_{0,+,-},T3, T4_{0,+,-},
T5_{0,+,-},$ and $T6$, are easily obtained \cite{KimKyaeSM}. The
representations are denoted as
\begin{equation} [{\bf SU(3)}_c,{\bf SU(2)}_W; {\bf SU(5)}',SU(3)']_{\tilde Y},
\end{equation}
where we already use the broken SU(3)$_W$ and $\tilde Y=Y_{\rm
Abel}+\frac1{\sqrt3}W_8$ given in Eq. (\ref{tildeY}). For obvious
cases, we will use the abbreviated notation
\begin{equation}
({\bf SU(3)}_c,{\bf SU(2)}_W)_{Y}.
\end{equation}
But when SU(3)$'$ triplets or antitriplets are involved, the
hypercharge is $\tilde Y$. We list all matter fields below,
\begin{align}
\begin{array}{l}
U_1:~ (\one,\two)_{1/2},\  2\cdot(\one,\two)_{-1/2},\
\one_1,\ 2\cdot\one_{0}\\
 U_2:~ (\one,\two)_{-1/2},\ \one_{0}\\
 U_3:~ (\one,\two)_{-1/2},\ 2\cdot(\one,\two)_{1/2},\
 2\cdot \one_{1}\\
T_{1_0}:~ (\threeb, \one)_{1/3},\ (\one,\two)_{1/2},\
 3\cdot\one_1,\ 2\cdot\one_0
 \\
 T_{1_-}:~  (\one;\fiveb',\one)_{0},\
  (\one;\one, \three')_{1/3},\ 2\cdot\one_{-1}
 \\
 T_{2_0}:~  (\threeb,\one)_{1/3},\
  (\one,\two)_{-1/2},\ 3\cdot\one_0
\\
T_{2_+}:~ (\one;\ten',\one)_{0},\ (\one;\one, \three')_{1/3},\
4\cdot\one_0
 \\
T_3:~ 2\cdot (\one;\five', \one)_{0},\
 2\cdot (\one;\fiveb', \one)_{0},\
 \\
\quad\quad (2_L+1_R)(\one,\two)_{1/2},\ (1_L+2_R)(\one,\two)_{1/2},\
\\
 \quad\quad (2_L+1_R)\one_{1},\ 3\cdot\one_0,\
   (6L+6R)\cdot\one_{1}
  \\
T_{4_0}:~ 3\cdot(\one,\two;\one, \threeb')_{1/6},\
 3\cdot( \one;\one, \threeb')_{-1/3} \\
T_{4_+}:~ 5\cdot(\one;\one, \three')_{1/3},\
 2\cdot(\one;\one, \three')_{-2/3}\\
T_{4_-}:~  3\cdot(\three,\two)_{1/6},\ 2\cdot(\threeb,\one)_{-2/3},\
\ 5\cdot(\threeb,\one)_{1/3},\ 3\cdot(\three,\one)_{-1/3},\
\\
 \quad\quad 5\cdot(\one,\two)_{-1/2},\
 2\cdot(\one,\two)_{1/2},\
2\cdot\one_{1},\ 12\cdot\one_{0},\  12\cdot\one_{0}\\
T_{7_0}:~ (\one;\fiveb',\one)_{0},\ (\one;\one,\three')_{-2/3} \\
T_{7_+}:~ (\threeb,\one)_{-2/3},\ (\three,\one)_{-1/3},\
2\cdot(\one,\two)_{-1/2} ,\ \one_0,\ 3\cdot\one_{-1} \\
T_{7_-}:~ (\one;\five',\one)_{0},\
(\one;\one,\threeb')_{-1/3},\ 2\cdot\one_1\\
T_6:~ 3\cdot(\one;\five',\one)_{0},\ 3\cdot(\one;\fiveb',\one)_{0},\
2\cdot(\one;\five',\one)_{1},\ 2\cdot(\one;\fiveb',\one)_{-1}
  \end{array}\label{Allspectrum}
 \end{align}
where $\one=(\one,\one,\one;\one,\one)$. Breaking SU(3)$'$, we
assign
\begin{align}
F_3=\textstyle (\frac12, -\frac12, 0) ,\quad
\frac1{\sqrt3}F_8=(\frac16, \frac16, -\frac13).
\end{align}
Then $\three'$ has extra entries of $\frac23, -\frac13, -\frac13$,
and $\threeb'$ has extra entries of $-\frac23, \frac13, \frac13$.
Thus, SU(3)$'$ (anti-)triplets of $T_{1_-}, T_{2_+}, T_{4_0},
T_{4_+}, T_{7_0}$ and $T_{7_-}$ are
\begin{align}
\begin{array}{l}
 T_{1_-}:~ (\one;\one, \three')_{1/3}\to \one_1,\ \one_0,\ \one_0
 \\
T_{2_+}:~  (\one;\one, \three')_{1/3}\to \one_1,\ \one_0,\ \one_0
 \\
T_{4_0}:~ 3\cdot(\one,\two;\one, \threeb')_{1/6} \to
3\cdot(\one,\two)_{-1/2},\ 3\cdot(\one,\two)_{1/2},\
 3\cdot(\one,\two)_{1/2},\\
 \quad\quad
 3\cdot( \one;\one, \threeb')_{-1/3}\to 3\cdot\one_{-1},\
 3\cdot\one_0,\ 3\cdot\one_0 \\
T_{4_+}:~ 5\cdot(\one;\one, \three')_{1/3}\to 5\cdot\one_1,\
5\cdot\one_0,\ 5\cdot\one_0\\
 \quad\quad 2\cdot(\one;\one, \three')_{-2/3}
 \to 2\cdot\one_0,\ 2\cdot\one_{-1},\ 2\cdot\one_{-1}\\
T_{7_0}:~  (\one;\one,\three')_{-2/3}\to \one_0,\ \one_{-1},\ \one_{-1} \\
T_{7_-}:~  (\one;\one,\threeb')_{-1/3}\to \one_{-1},\ \one_0,\
\one_0
  \end{array}\label{SU3spec}
 \end{align}
Eq. (\ref{Allspectrum}) with (\ref{SU3spec}) gives the SM quantum
numbers. From these, we note that there is {\it no exotics}. Other
exotics free orbifold compactifications \cite{KimKyaeSM,Kim07} have
\Eh\ sector contribution to $Y$ as in the present case. But, we do
not know whether this is a necessary condition for exotics free
models or not.

 \begin{table}[t]
\begin{center}
\begin{tabular}{|c|c|c|c|c|}
\hline  $P+[kV+ka]$  &
No.$\times$(Repts.)$_{Y[Q_1,Q_2,Q_3,Q_4,Q_5]}$
 &$\Gamma$ &Label\\
\hline
 $(\underline{\frac{-1}{3}~\frac{-1}{3}~\frac{-2}{3}}~
 \underline{\frac{2}{3}~\frac{-1}{3}}
 ~\frac{-1}{3}~0~0)(0^8)'_{T_{4_-}}$
  & $
 3\cdot(\three,\two)_{1/6~[0,0,0;0,0]}^L$
  & $1$ &$ q_1,~ q_2,~  q_3$
\\
$(\underline{\frac16~\frac{1}{6}~\frac{5}{6}}~
 {\frac{1}{6}~\frac{1}{6}}
 ~\frac{1}{6}~\frac12~\frac12)(0^8)'_{T_{4_-}}$   &$
 2\cdot(\threeb,\one)_{-2/3~[-3,3,2;0,0]}^L$
  & $3$ &$ u^c,~  c^c$
\\
$(\underline{\frac{-1}{3}~\frac{-1}{3}~\frac{-2}{3}}~
 {\frac{1}{3}~\frac{1}{3}}~\frac{1}{3}
 ~\frac{-1}{4}~\frac{-1}{4})(\frac14 ^5~\frac{1}{12}~\frac{1}{12}
 ~\frac{1}{12})'_{T_{7_+}}$   &$
 (\threeb,\one)_{-2/3~[0,6,-1;5,1]}^L$
  & $1$ &$ t^c$
\\
$(\underline{\frac{1}{2}~\frac{1}{2}~\frac{1}{2}}~
 {\frac{-1}{6}~\frac{-1}{6}}~\frac{-1}{6}
 ~0~0)(0^5~\frac{-1}{3}~\frac{-1}{3}
 ~\frac{-1}{3})'_{T_{2_0}}$   &$
 (\threeb,\one)_{1/3~[3,-3,0;0,-4]}^L$
 & $-1$ &$ d^c$
\\
$(\underline{\frac16~\frac{1}{6}~\frac{5}{6}}~
 {\frac{1}{6}~\frac{1}{6}}
 ~\frac{1}{6}~\frac{-1}2~\frac{-1}2)(0^8)'_{T_{4_-}}$  &$
 2\cdot(\threeb,\one)_{1/3~[-3,3,-2;0,0]}^L$
 & $1$ &$ s^c,~ b^c$
\\
[0.2em]\hline
 $({\frac{-1}{3}~\frac{-1}{3}~\frac{1}{3}}~
 \underline{\frac{2}{3}~\frac{-1}{3}}~\frac{2}{3}
 ~0~0)(0^8)'_{T_{4_-}}$  &$
 (\one,\two)_{-1/2~[-6,6,0;0,0]}^L$
  & $1$ &$l_1,l_2,l_3$
\\
[0.2em]\hline
 $(0~0~0~\underline{\frac{2}{3}~\frac{-1}{3}}~
 \frac{2}{3}~\frac{-1}{4}~\frac{-1}{4})
 (\frac14 ^5~\frac{1}{12}~\frac{1}{12}
 ~\frac{1}{12})'_{T_{1_0}}$  &$
 (\one,\two)_{1/2~[0,6,-1;5,1]}^L$
  & $0$ &$H_u$
\\
$(\frac{-1}{3}~\frac{-1}{3}~\frac{1}{3}~
\underline{\frac{1}{3}~\frac{-2}{3}}~\frac{1}{3}~
  ~\frac{-1}{4}~\frac{-1}{4})(\frac14 ^5~\frac{1}{12}~\frac{1}{12}
 ~\frac{1}{12})'_{T_{7_+}}$  &$
 (\one,\two)_{-1/2~[-6,0,-1;5,1]}^L$
  & $-2$ &$H_d$
\\
[0.2em]\hline
\end{tabular}
\end{center}
\caption{Three families of quarks and leptons and a pair of Higgs
doublets. We do not list singlet leptons since there are many
possibilities.} \label{table:Families}
\end{table}

\subsection{Three families with no exotics}

Removing vectorlike representations and neutral singlets, we obtain
the following chiral representations,
\begin{align}
&T_{4_-, 7_+, 1_0}:~  3\cdot(\three,\two)_{1/6},~
3\cdot(\three,\one)_{-2/3},~ 3\cdot(\three,\one)_{1/3},~
  3\cdot(\one,\two)_{-1/2},~ 3\cdot\one_1 \label{SMdoublet}
\\
 &\quad T_{2_+,7_0}:~ \ten'_{0},~ \fiveb'_{0}
\end{align}
where $\ten'_{0}=(\one;\ten',\one)_{0}$ and
$\fiveb'_{0}=(\one;\fiveb',\one)_{0}$. In Table
\ref{table:Families}, we list three families except the charged
lepton singlets. Note that SU(3)$_c$ triplets with underlined
entries mean, for example,
$(\underline{\frac{-1}{3}~\frac{-1}{3}~\frac{-2}{3}})=
({\frac{-1}{3}~\frac{-1}{3}~\frac{-2}{3}}),
({\frac{2}{3}~\frac{-1}{3}~\frac{1}{3}}),
({\frac{-1}{3}~\frac{2}{3}~\frac{1}{3}})$, and
$(\underline{\frac{1}{6}~\frac{1}{6}~\frac{5}{6}})=
({\frac{1}{6}~\frac{1}{6}~\frac{5}{6}}),
({\frac{1}{6}~\frac{-5}{6}~\frac{-1}{6}}),
({\frac{-5}{6}~\frac{1}{6}~\frac{-1}{6}})$. This is because of the
asymmetrical simple roots of SU(3)$_c$ in Eq. (\ref{SU3croots}).

\subsection{Matter parity}\label{subsec:Parity}

Let us define the U(1)$_\Gamma$ charge as a linear combination of
$Q_{1-5}$ of Eq. (\ref{U1s}) and $W_8$. We choose its generator
$\Gamma$ such that the light quarks carry odd U(1)$_\Gamma$ charges
while Higgs doublets carry even  U(1)$_\Gamma$  charges. This is
necessary to remove the baryon number violating $u^cd^cd^c$ term.
For the lepton number violation, the condition is not so strong and
furthermore in our model there are so many possibilities in choosing
the charged singlets $e^c$, and here we do not discuss them. Then,
one successful choice of $\Gamma$ is
\begin{equation}
\Gamma=\frac13 Q_2+ Q_3+\tilde W_8 \label{UGamma}
\end{equation}
where
$$\tilde W_8=(0^3~1~1~\textstyle{-2}~0^2)(0^8)'.$$
 The $\Gamma$ quantum numbers are also listed in Table
 \ref{table:Families}. Breaking
U(1)$_\Gamma$ by VEVs of even integer SM singlets, a discrete
symmetry $\Z_2$, which is called {\it matter parity} $P$, survives,
\begin{equation}
U(1)_\Gamma\to P.\label{Rparity}
\end{equation}
Thus, looking at the light quarks only the dangerous term
$u^cd^cd^c$ is not allowed. However, we have to consider mixing of
light quarks with heavy quarks which can be dangerous in principle
\cite{KimKyaeSM}. In our model, there are ten quark flavors: six SM
quarks and four extra \Qem$=-\frac13$ quarks denoted as $3\cdot
(D+\overline{D})$ and $(D'+\overline{D}')$. For quark mixing, we
need to consider $\overline{D}$s and $\overline{D}'$ only. In Eq.
(\ref{Allspectrum}), three $\overline{D}$s (three out of five
$(\threeb,\one)_{1/3}$s) in $T_{4_-}$ appear as
$(\threeb,\one)_{1/3~[6,-6,0;0,0]}$ carrying $\Gamma=-2$ and
$\overline{D}'$ in $T_{1_0}$ appears as
$(\threeb,\one)_{1/3~[3,3,1;5,1]}$ carrying $\Gamma=2$. Therefore,
if $P$ is not broken, light $d^c$ and heavy $\overline{D}$s and
$\overline{D}'$ can never mix and we achieve an exact matter parity
$P$. But a successful matter parity assignment should not be in
conflict with other phenomenological requirements. The most severe
constraint comes from making exotic particles massive
\cite{KimKyaeSM}. In passing, we point out that the other vectorlike
particles, such as $D-\overline{D}, D'-\overline{D}'$, doublet
pairs, and unit charge lepton pairs $E^--\overline{E}^+$, are not so
dangerous as exotics. Since our model does not include any exotics,
we do not need VEVs of any odd $\Gamma$ singlets for obvious
phenomenological reasons. A detailed study of singlet VEVs is
outside of the scope of the present discussion, and will be
presented elsewhere.

 \begin{table}[t]
\begin{center}
\begin{tabular}{|c|c|c|}
\hline  $P+n[V\pm a]$ &  $\Gamma$ &
No.$\times$(Repts.)$_{Y[Q_1,Q_2,Q_3,Q_4,Q_5]}$
\\
\hline
$({\frac{1}2~\frac{1}2~\frac{-1}2~\underline{\frac{1}{2}~\frac{-1}{2}}~
 \frac{1}{2}~\frac{-1}{2}~\frac{-1}{2}})
 (0^8)'_{U_1}$
 & $-2$ & $
 (\one,\two)_{\frac12~[9,3,-2;0,0]}^L$
\\
$(0~0~0~\underline{1~0}~ 0~\underline{1~0})
 (0^8)'_{U_1}$
 & $4$ & $
 (\one,\two,\two)_{-\frac12~[0,6,2;0,0]}^L$
\\
$({0~0~0~\underline{1~0}~ 1~0~0})
 (0^8)'_{U_2}$
 & $3$ & $
 (\one,\two)_{-\frac12~[0,12,0;0,0]}^L$
\\
$({\frac12~\frac12~\frac{-1}2~\underline{\frac12~\frac{-1}2}~
\frac12~\frac12~\frac12})
 (0^8)'_{U_3}$
 & $2$ & $
 (\one,\two)_{-\frac12~[9,3,2;0,0]}^L$
\\
$(0~0~0~\underline{0~\textstyle{-1}}~
0~\underline{0~\textstyle{-1}})
 (0^8)'_{U_3}$
 & $-4$ & $
 (\one,\two,\two)_{\frac12~[0,-6,-2;0,0]}^L$
\\
$(0~0~0~\underline{\frac23~\frac{-1}3}~
\frac23~\frac{-1}4~\frac{-1}4)
 (\frac{1}4~\frac{1}4~\frac{1}4~\frac{1}4~\frac{1}4~\frac1{12}
 ~\frac1{12}~\frac1{12})'_{T_{1_0}}$
 & $0$ & $
 \star\ (\one,\two)_{\frac12~[0,6,-1;5,1]}^L$
\\
$(0~0~0~\underline{\frac13~\frac{-2}3}~ \frac13~\frac{1}2~\frac{1}2)
 (0^5~\frac{-1}3
 ~\frac{-1}3~\frac{-1}3)'_{T_{2_0}}$
 & $1$ & $
 (\one,\two)_{-\frac12~[0,0,2;0,-4]}^L$
\\
$(0~0~0~\underline{0~\textstyle{-1}}~ 0~\frac{1}4~\frac{1}4)
 (\frac{1}4~\frac{1}4~\frac{1}4~\frac{1}4~\frac{1}4~\frac{-1}4
 ~\frac{-1}4~\frac{-1}4)'_{T_{3}}$
 & $-2$ & $
 (2_L+1_R)\cdot(\one,\two)_{\frac12~[0,-6,1;5,-3]}^L$
\\
$(0~0~0~\underline{0~\textstyle{-1}}~ 0~\frac{-1}4~\frac{-1}4)
 (\frac{-1}4~\frac{-1}4~\frac{-1}4~\frac{-1}4~\frac{-1}4~\frac{1}4
 ~\frac{1}4~\frac{1}4)'_{T_{3}}$
 & $-4$ & $
 (2_L+1_R)\cdot(\one,\two)_{-\frac12~[0,-6,-1;-5,3]}^L$
\\
 $(0~0~0~\underline{\frac{2}{3}~\frac{-1}{3}}~
 \frac{-1}{3}~0~0)
 (0^5~\underline{\frac{-2}{3}~\frac{1}{3}~
 \frac{1}{3}})'_{T_{4_0}}$
 & $1$ & $
 6\cdot(\one,\two)_{\frac12~[0,0,0;0,0]}^L$
\\
$(\frac{-1}3~\frac{-1}3~\frac{1}3~\underline{\frac{2}3~\frac{-1}3}~
\frac{2}3~0~0)
 (0^8)'_{T_{4_-}}$
 & $1$ & $
\star\ 3\cdot(\one,\two)_{-\frac12~[-6,6,0;0,0]}^L$
\\
$(\frac{1}6~\frac{1}6~\frac{-1}6~\underline{\frac{1}6~\frac{-5}6}~
\frac{1}6~\frac12~\frac12)
 (0^8)'_{T_{4_-}}$
 & $0$ & $
 2\cdot(\one,\two)_{-\frac12~[3,-3,2;0,0]}^L$
\\
$(\frac{1}6~\frac{1}6~\frac{-1}6~\underline{\frac{1}6~\frac{-5}6}~
\frac{1}6~\frac{-1}2~\frac{-1}2)
 (0^8)'_{T_{4_-}}$
 & $-4$ & $
 2\cdot(\one,\two)_{\frac12~[3,-3,-2;0,0]}^L$
\\

$(\frac{-1}3~\frac{-1}3~\frac{1}3~\underline{\frac13~\frac{-2}3}~
\frac13~\frac{-1}4~\frac{-1}4)
 (\frac{1}4~\frac{1}4~\frac{1}4~\frac{1}4~\frac{1}4~\frac1{12}
 ~\frac1{12}~\frac1{12})'_{T_{7_+}}$
 & $-2$ & $
 (\one,\two)_{-\frac12~[-6,0,-1;5,1]}^L$
\\
$(\frac{1}6~\frac{1}6~\frac{-1}6~\underline{\frac{5}6~\frac{-1}6}~
\frac{-1}6~\frac{1}4~\frac{1}4)
 (\frac{1}4~\frac{1}4~\frac{1}4~\frac{1}4~\frac{1}4~\frac1{12}
 ~\frac1{12}~\frac1{12})'_{T_{7_+}}$
 & $3$ & $
 (\one,\two)_{-\frac12~[3,3,1;5,1]}^L$
\\
[0.2em] \hline $(0~0~0~\underline{\frac{2}{3}~\frac{-1}{3}}~
 \frac{-1}{3}~0~0)
 (0^5~\underline{\frac{-2}{3}~\frac{1}{3}~
 \frac{1}{3}})'_{T_{4_0}}$
 & $1$ & $
 3\cdot(\one,\two)_{\frac{-1}2~[0,0,0;0,0]}^L$
\\[0.2em]
\hline
\end{tabular}
\end{center}
\caption{Thirty-three color-singlet SU(2)$_W$ doublets which contain
the leptons (the last row) and Higgs particles. The MSSM pair is
starred.} \label{table:doublets}
\end{table}

\subsection{Higgs doublets}\label{subsec:HiggsD}

In Table \ref{table:doublets}, we list all color singlet doublets,
where we included lepton doublets in the last row. Higgs doublets
form a vectorlike representation under the SM gauge group. So, they
can be removed at the GUT scale in principle. One vectorlike pair
$H_u+H_d$ is kept light for breaking the SU(2)$\times$U(1)$_Y$ gauge
symmetry at the electroweak scale. We choose the starred doublets to
give large masses to $t$ and $b$ quarks. We choose $H_u$ such that
the sum of the sector numbers in $q_3t^c H_u$ adds up to 0 mod 12.
Then, $H_u$ is chosen from $T_{1_0}$. For $b$ quark, a similar
argument chooses one $(\one,\two)_{-1/2}$ in $T_{4_-}$ as $H_d$.
These $H_u$ and $H_d$ are starred in Table \ref{table:doublets}.
However, note that this is just one illustration and another choice
may well be possible depending on the Yukawa couplings and
magnitudes of singlet VEVs.

\subsection{The Lee-Weinberg model}

This model is basically a string realization of the Lee-Weinberg
model based on SU(3)$_c\times$SU(3)$_W\times$U(1) \cite{LW77}. In
the Lee-Weinberg model, one quark family consists of
\begin{equation}
\three_{W,q}=\left( \begin{array}{ccc}
d & & u\\
 & D &
\end{array}
\right)_L, \quad d_R  , u_R,
 D_R\label{LWq}
\end{equation}
Thus, our model realizes just three left-handed quark triplets with
{\it no extra $\three_W-\threeb_W$ quark pairs}, and hence it is a
minimal kind of Lee-Weinberg type models. Out of 21 left-handed
$\three_W$s and 21 left-handed $\threeb_W$s, 12 pairs form a
vectorlike representations under the Lee-Weinberg gauge
group.\footnote{The breaking scale of SU(3)$_W$ can be very low in
principle, because the discrepancy in the numbers of multiplets
between SU(3)$_c$ (ten flavors) and SU(3)$_W$ (twenty-one flavors)
enables one to lower the breaking scale of SU(3)$_W$ while
generating the difference of gauge couplings of SU(3)$_c$ and
SU(3)$_W$. But, we will not consider this possibility here.} This is
gleaned from the chiral representation (\ref{SMdoublet}) that there
remain three pairs of $(\three_c,\three_W)$. Thus, for SU(3)$_W$
anomaly cancellation, there must be nine $\threeb_W$s, and the
remaining $\three_W-\threeb_W$ pairs must form a vectorlike
representation. [We include the odd $\Gamma$ Higgs pairs of Table
\ref{table:doublets} in the vectorlike representation.] Nine
color-singlet $\threeb_W$s contain three lepton doublets and three
pairs of Higgs doublets. The electromagnetic charges of nine
$\threeb_W$s contain three $\threeb_{W,+}$ and six $\threeb_{W,0}$,
where
\begin{equation}
\bar\three_{W,+}=\left( \begin{array}{ccc}
 &\psi_1^+ & \\
 \psi^0&  &\psi_2^+
\end{array}
\right)_L,\quad
 \bar\three_{W,0}=\left( \begin{array}{ccc}
 &\psi_1^0 & \\
 \psi^-&  &\psi_2^0
\end{array}
\right)_L\label{LWlH}
\end{equation}
where $\psi^{\rm sign}$ denotes the integer electromagnetic charge
of the field $\psi$. In Eqs. (\ref{LWq}) and (\ref{LWlH}), SU(2)$_W$
doublets are pairs of $u-d, \psi_2^+-\psi_0,$ and $\psi_2^0-\psi^-$.
Obviously, three lepton doublets of (\ref{SMdoublet}) must come from
three $\bar\three_{W,0}$s, and we are left with three pairs of
$\bar\three_{W,0}-\bar\three_{W,+}$.

\subsection{The $\mu$ term}

A possible large $\mu$ term arises from the coupling between three
pairs of $\bar\three_{W,0}-\bar\three_{W,+}$ as
$\epsilon_{\alpha\beta\gamma}\threeb_W^\alpha
\threeb_W^\beta\threeb_W^\gamma$ where $\alpha,\beta,\gamma$ are
SU(3)$_W$ indices. Suppose that SU(3)$_W$ is broken by VEVs
(typically of order $V$) of $\psi_1^0$ in $ \bar\three_{W,0}$ (and
also by $\three_{W,0}$ in the removed vectorlike representation
toward a $D$-flat condition). Then, the $H_u-H_d$ type couplings
arise from\footnote{Note that $\three_W-\threeb_W$ coupling is not
generating $H_u-H_d$ terms since both $H_u$ and $H_d$ belong to
$\threeb_W$.}
\begin{equation}
\epsilon_{\alpha\beta\gamma}\threeb_{W,I}^\alpha
\threeb_{W,J}^\beta\threeb_{W,K}^\gamma\epsilon^{IJK}\sim V
\epsilon_{\alpha\beta}\threeb_{W,I}^\alpha
\threeb_{W,J}^\beta\epsilon^{IJ}\label{antisymmY}
\end{equation}
where  $I,J,K$ are the Higgs family indices. For a general family
coupling $g^{IJK}$, due to $\epsilon_{\alpha\beta\gamma}$ the
symmetric part does not give an $H_u-H_d$ coupling because it gives,
$\propto \threeb_{W,\bar 1}\threeb_{W,\bar 2}- \threeb_{W,\bar
2}\threeb_{W,\bar 1} =0$. Because of $\epsilon^{IJ}$, the same Higgs
family does not have the coupling and the $3\times 3$ $H_u-H_d$ mass
matrix is an antisymmetric one whose determinant is zero. Therefore,
we obtain two massive Higgs doublet pairs and one massless Higgs
doublet pair. Thus, there remains only one massless Higgs doublet
pair, achieving the MSSM spectrum at low energy. In this scheme
also, there are methods to generate an electroweak scale $\mu$ term
\cite{Mu,MuSol}.

\section{Hidden sector SU(5)$'$, gauge mediation and messengers}
\label{sec:SU(5)}

As shown in Table \ref{table:Hidden}, there are SU(5)$'$ fields. But
some of these obtain masses by Yukawa couplings at the string scale.
Below the string scale vectorlike pairs become massive by VEVs of
singlets, and hence we consider only the chiral representations. We
need the mass scale of the vectorlike pairs are much above the
SU(5)$'$ confining scale so that the SUSY breaking by $\ten'$ and
$\fiveb'$ is intact.

 \begin{table}[t]
\begin{center}
\begin{tabular}{|c|c|c|}
\hline  $P+n[V\pm a]$ &  $\Gamma$ &
No.$\times$(Repts.)$_{Y[Q_1,Q_2,Q_3,Q_4,Q_5]}$
\\
\hline $({\frac{1}6~\frac{1}6~\frac{-1}6~\frac{1}{6}~\frac{1}{6}~
 \frac{1}{6}~\frac{1}{4}~\frac{1}{4}})
 (\underline{\frac{-3}4~\frac{1}{4}~\frac{1}{4}~\frac{1}{4}
 ~\frac{1}{4}}~
 \frac{-1}{4}~\frac{-1}{4}~\frac{-1}{4})'_{T1_-}$
 & $2$ & $
 (\one;\fiveb', \one)_{0~[3,3,1;1,-1]}^L$
\\
$(\frac{1}{6}~\frac{1}{6}~\frac{-1}{6}~~\frac{-1}{6}~\frac{-1}{6}~
 \frac{-1}{6}~0~0)
 (\underline{\frac{1}{2}~\frac{1}{2}~\frac{-1}{2}~\frac{-1}{2}
 ~\frac{-1}{2}~}\frac{-1}{6}~\frac{-1}{6}~\frac{-1}{6}~)'_{T2_+}$
  & $-1$ & $\star~({\bf 1};\ten', \one)_{0~[3,-3,0;-2,-2]}^L$
\\
 $(0^6~\underline{\frac14~\frac{-3}{4}})
 (\underline{\frac34~\frac{-1}{4}~\frac{-1}{4}~\frac{-1}{4}
 ~\frac{-1}{4}}~\frac{1}{4}~\frac{1}{4}
 ~\frac{1}{4})'_{T3}$
  & $-1$ & $(\two_n;{\bf 5}',\one)_{0~[0,0,-1;-1,3]}^L$
\\
  $(0^6~\underline{\frac34~\frac{-1}{4}})
 (\underline{\frac{-3}4~\frac{1}{4}~\frac{1}{4}~\frac{1}{4}
 ~\frac{1}{4}}~\frac{-1}{4}~\frac{-1}{4}
 ~\frac{-1}{4})'_{T9}$
  & $1$ & $(\two_n;\fiveb',\one)_{0~[0,0,1;1,-3]}^L$
  \\
 $(0^3~{\frac{-1}{3}~\frac{-1}{3}}~\frac{-1}{3}~\frac{1}{4}~\frac{1}{4})
 (\underline{\frac{-3}4~\frac{1}{4}~\frac{1}{4}~\frac{1}{4}
 ~\frac{1}{4}}~\frac{1}{12}~\frac{1}{12}
 ~\frac{1}{12})'_{T7_0}$
  & $-1$ & $\star~(\one;\fiveb',\one)_{0~[0,-6,1;1,1]}^L$
\\
 $(\frac{1}6~\frac{1}6~\frac{-1}6~
 \frac{1}6~\frac{1}6~\frac{1}6~\frac{-1}{4}~\frac{-1}{4})
 (\underline{\frac{3}4~\frac{-1}{4}~\frac{-1}{4}~\frac{-1}{4}
 ~\frac{-1}{4}}~\frac{1}{4}~\frac{1}{4}~\frac{1}{4})'_{T7_-}$
  & $0$ & $(\one;\five',\one)_{0~[3,3,-1;-1,3]}^L$
\\
$(0^6~\frac{-1}{2}~\frac{-1}{2})
 (\underline{\textstyle -1~0~0~0~0}~0~0~0)'_{T6}$ & $-2$ & $
 3\cdot(\one;\fiveb', \one)_{0~[0,0,-2;-4,0]}^L$
\\
$(0^6~\frac{-1}{2}~\frac{-1}{2})
 (\underline{\textstyle 1~0~0~0~0}~0~0~0)'_{T6}$ & $-2$ & $
 2\cdot(\one;{\bf 5}', \one)_{1~[0,0,-2;4,0]}^L$
\\
$(0^6~\frac{1}{2}~\frac{1}{2})
 (\underline{\textstyle -1~0~0~0~0}~0~0~0)'_{T6}$ & $2$ & $
 2\cdot(\one;\fiveb', \one)_{-1~[0,0,2;-4,0]}^L$
\\
$(0^6~\frac{1}{2}~\frac{1}{2})
 (\underline{\textstyle 1~0~0~0~0}~0~0~0)'_{T6}$ & $2$ & $
 3\cdot(\one;\five,\one)_{0~[0,0,2;4,0]}^L$
\\[0.2em]
\hline
\end{tabular}
\end{center}
\caption{Hidden sector SU(5)$'$  representations under
SU(2)$_n\times $SU(5)$'\times $SU(3)$'$. After removing vectorlike
representations by $\Gamma=$ even integer singlets, the starred
representations remain.} \label{table:Hidden}
\end{table}

In Table \ref{table:Hidden}, we list all the SU(5)$'$ non-singlet
fields. From these, one can easily check that SU(5)$'$ gauge anomaly
is absent. One conspicuous feature is that we obtained one $\ten'$.
Except $\ten'$ of $T_{2_-}$ and $\fiveb'$ of $T_{7_0}$, the rest 8
flavors form a vectorlike representation under
SU(5)$'\times$SU(2)$_n\times$U(1)$_Y$. Removal of the eight flavors
much above the SU(5)$'$ confining scale is achieved by VEVs of SM
gauge singlet fields, breaking extra gauge symmetries. It has been
known that $\ten'+\fiveb'$ of a confining SU(5)$'$ breaks SUSY
\cite{ADS} and {\it we achieve the GMSB if the confining scale is
below $10^{12}$ GeV.} Note that $\ten'_0$ and $\fiveb'_0$ do not
carry any SU(3)$_c\times$SU(2)$_W\times$U(1)$_Y$ charge (which is
emphasized by the subscript 0) and DSB by $\ten'_0$ and $\fiveb'_0$
does not break the SM gauge group.

Note that the singlet combination $\ten'\ten'\ten'\fiveb'$ is not
possible with one $\ten'$. The SU(5)$'$ singlet combination in this
uncalculable model can be parameterized by the gauge field strength
field ${\cal W'}^\alpha{\cal W}'_\alpha$ as discussed in
\cite{Murayama07}.  The interaction
between the messenger $f$ and the hidden sector gauge fields
can appear from string compactification as
\begin{equation}
{\cal L}=\int d^2\theta \left[\xi(S_1,S_2,\cdots) f\bar{f}
 {\cal W'}^\alpha{\cal W'}_\alpha+\eta(S_1,S_2,\cdots)
 f\overline{f}\right]+{\rm h.c.}\label{yukawast}
\end{equation}
where we have in general the holomorphic functions $\xi$ and
$\eta$ of singlet chiral fields, $S_1, S_2,\cdots$. The quantum
number of $\xi(S_1,S_2,\cdots) f\bar{f}$ is the same as that of
dilaton, where the $H$-momentum of dilaton is $(0,0,0)$.
On the other hand, the $H$-momenta
of the superpotential term $\eta(S_1,S_2,\cdots)
f\overline{f}$ should be $(-1,1,1)$.
The $H$-momenta of the twisted sectors are given by
\cite{Katsuki,KimKyaeSM,ChoiKimBk}
\begin{align}
&U_1: (-1,0,0),\quad U_2: (0,1,0),\quad U_3:
(0,0,1),\nonumber\\
&\textstyle T_1:(\frac{-7}{12},\frac{4}{12},\frac1{12}),\quad
 T_2:(\frac{-1}{6},\frac46,\frac16),\quad T_3:
 (\frac{-3}{4},0,\frac{1}{4}),\nonumber\\
&\textstyle
 T_4:(\frac{-1}{3},\frac13,\frac13),\quad
\left\{T_5:(\frac{1}{12},\frac{-4}{12},\frac{-7}{12})\right\}, \quad
T_6:(\frac{-1}{2},0,\frac12),\\
&\textstyle T_7:(\frac{-1}{12},\frac{4}{12},\frac{7}{12}),\quad
T_9:(\frac{-1}{4},0,\frac{3}{4}) , \nonumber
\end{align}
The Yukawa coupling
$\eta(S_1,S_2,\cdots) f\overline{f}$ and the coefficient of
${\cal W'}^\alpha{\cal W'}_\alpha$ must satisfy the modular invariance
rule for the twisted sector fields($z$) multiplication,
\begin{equation}
\sum_z k(z)=0\ {\rm mod}\ 12,\quad \sum_z [km_f](z)=0\ {\rm mod}\ 3.
\label{modrule}
\end{equation}
Consider, for example, the vectorlike colored particles appearing
only in $T_{4_-}$ with \Qem$=\mp\frac13$: $f_3=D, \bar
f_3=\overline{D}$, viz. Eq. (\ref{Allspectrum}).
We can consider the following gauge singlet combination multiplied
to ${\cal W'}^\alpha{\cal W'}_\alpha$, for example,
\begin{equation}
T_{4_-}T_{4_-}T_{1_0}T_{7_+}T_{4_-}T_{4_-}\sim \bar f_3f_3\langle
T_{1_0}T_{7_+}T_{4_-}T_{4_-}\rangle\sim D_{-1/3}\overline{D}_{1/3}.
 \label{mindim}
\end{equation}
Similarly, SU(2)$_W$ doublet coupling
${\cal W'}^\alpha{\cal W'}_\alpha'$
can be considered. The product in
(\ref{mindim}), $T_{4_-}T_{4_-}T_{1_0}T_{7_+}T_{4_-}T_{4_-}$,
has the $H$-momentum $(-2, 2, 2)$, and hence we must
multiply further singlets to make the sum of
$H$-momenta be $(0, 0, 0)$. As shown in
\cite{KimKyaeSM}, usually we can achieve this, but here we do not
elaborate the details. In this model, $f_3$ and $f_2$ denote the messenger
through SU(3)$_c$ and the messenger through SU(2)$_W$, respectively.
If needed, we can also consider $f_1$ (the messenger through U(1)$_Y$).
Below, $f$ represents $f_3,f_2,$ or $f_1$.

From the above discussion, the
fields of $f,\bar f$ and ${\cal W'}^\alpha{\cal W'}_\alpha$ can
have the following tree level Lagrangian,
\begin{equation}
{\cal L}=\int d^2\theta \left[\frac{1}{M^2} f\bar{f}
 {\cal W'}^\alpha{\cal W'}_\alpha+M_f
 f\overline{f}\right]+{\rm h.c.}\label{Mura}
 \end{equation}
which is perturbative in origin.
Here $M$ and $M_f$ are determined by the strength of coupling constant
and VEVs of singlet fields appearing in $\xi$ and $\eta$ of Eq.
(\ref{yukawast}). Both of these parameters are assumed to be
somewhat less than the string scale. The SUSY breaking through Eq.
(\ref{Mura}) has been discussed in \cite{Murayama07} by introducing
the messenger mass- and F-parameters
\begin{equation}
M_{\rm mess}\thickapprox M_f+\frac{\Lambda_h^3}{M^2},
\quad F_{\rm mess}\thickapprox \frac{\Lambda_h^4}{M^2}.
\end{equation}
With this GMSB scenario, firstly the observable sector
gaugino obtains mass of order
\begin{equation}
\tilde m_{\rm SUSY}\sim
\frac{g^2}{16\pi^2}\frac{\Lambda_h^4}{M^2M_{\rm mess}}
\end{equation}
while the gravitino mass is around $m_{3/2}\thickapprox
\Lambda_h^3/M_{Pl}^2$. To obtain
1 TeV gluino mass (but much smaller gravitino mass of order 0.2 GeV)
with  $\alpha=\frac1{25}$ and $\Lambda_h=10^{12}$ GeV,
for example, we need $(M^2 M_{\rm mess})^{1/3}\thickapprox 1.5\times
10^{14}$ GeV.

This leads us to consider the ${\cal W'}^\alpha{\cal W'}_\alpha$
couplings to $H_uH_d$ and the observable sector Yukawa couplings $W_Y
\sim H_u q_i u_j^c+ H_d q_i d_j^c$. Let us focus on the $H_uH_d$
coupling. From the discussion with (\ref{antisymmY}), the three pairs of
Higgsinos form an antisymmetric mass matrix parametrized by $A,B$ and $C$
which are assumed to be large.
The $\int d^2\theta H_uH_d{\cal W'}^\alpha{\cal W'}_\alpha$ term would
contribute to the Higgsino mass matrix and also to the soft $B$ parameter
matrix. The heavy pairs of $H_u$ and $H_d$ act as $f_2$ and
$\bar{f}_2$. We are interested in the light $H_u$ and $H_d$ pair.
The Higgsino mass matrix and the $B$ matrix take the following form,
\begin{align}
&M_{\rm Higgsino}=\left(\begin{array}{ccc} 0,& A+a, & B+b\\
   -A-a,& 0,  & C+c\\
  -B-b, & -C-c,  &0
 \end{array} \right)\label{MHiggsino} \\
&B_{\rm soft}=\mu\left(\begin{array}{ccc} 0,& a, & b \\
  -a, &  0, & c \\
   -b, & -c,  &0
 \end{array} \right)\label{MBpara}
\end{align}
where the parameters $a,b,c$ in (\ref{MHiggsino}) get contribution
from the hidden-sector gluino condensation while $\mu(a,b,c)$ in (\ref{MBpara})
get contribution from the $F$-term of ${\cal W'}^\alpha{\cal W'}_\alpha$. If $a:b:c=
A:B:C$, then the light Higgsinos and light Higgs bosons are paired to
constitute the Higgs multiplets of the MSSM. This proportionality can be achieved
if the same singlet combination is multiplied to the six nonvanishing superpotential
terms implied in (\ref{MHiggsino}) comprised of the
$H$-momentum $(-1,1,1)$ to make the $H$-momentum $(0,0,0)$ of $\xi f\bar f$
in (\ref{yukawast}). One may choose a vacuum so that such a condition is satisfied.
The interaction  $\int d^2\theta (\frac{1}{m^3}H_uq u^c+\cdots)
{\cal W'}^\alpha{\cal W'}_\alpha$ can be within a safe region of the
gauge hierarchy solution. For example, the $A$-term estimated from this is
$
A\thickapprox\frac{\Lambda_h^4}{m^3}
$
which can be of order $10^{-2}$ GeV -- $10^{6}$ GeV for
$\Lambda_h\sim 10^{10-12}$ GeV and $m\sim 10^{14}$ GeV.

Finally, we comment on possible higher order terms in the K\"ahler potential. Even though
all the important hidden sector matter $\ten'$ does not appear in the superpotential,
it can appear in the K\"ahler potential. Possible terms of the form
$\ten'\ten^{'*}ff^*/M_K^2$ might appear. The higher order K\"ahler terms was
calculated for the compactification $T^6=(T_2)^3$ (with the volume moduli $T$s and
the complex structure moduli $C$s) in Ref. \cite{Bailin} for
two matter fields $Q_\alpha$,
$$
K^{\rm matter}= \prod_{i=1}^3(T_i+\overline{T}_i)^{n_\alpha^i}
\prod^{h_{(2,1)}}_{m=1}(C_m+\overline{C}_m)^{l_\alpha} |Q_\alpha|^2
$$
where $n_\alpha^i$ and $h_{2,1}=1$ (for our $\Z_{12-I}$) are the modular weight and a Hodge number, respectively. Also, $l_\alpha$ is an integer. The term $\ten'\ten'^{*}ff^*/M_K^2$ is not appearing
in the above expression, and at present there does not exist a $K^{\rm matter}$
calculation for four matter fields of our interest.
Even if it appears, the mass suppression scale $M_K$ is expected to be of
order the string scale and hence is much larger than $M$ appearing in Eq. (\ref{Mura})
toward the GMSB scenario. However, if it appears with the same order of
the suppression factor as  in Eq. (\ref{Mura}), the idea of our GMSB is not successful phenomenologically. We may need $M^2/M_K^2<0.03$ \cite{DNShir}.

\section{Conclusion}

We have shown that there exists a possibility of the hidden sector
SU(5)$'$ with $\ten'_0$ plus $\fiveb'_0$ matter below the GUT scale
so that a GMSB at the stable vacuum is successful. Toward achieving
the needed coupling constant $\alpha'_5$ of the hidden sector at the
GUT scale, we may need different compactification radii for the
three tori \cite{Kim07}. The model is very interesting in that it
contains three MSSM families without any exotics. We find a
desirable U(1)$_\Gamma$ gauge symmetry whose $\Z_2$ discrete group
can be a matter parity $P$ or $R$-parity. Due to our Lee-Weinberg
type model, there remains only one light pair of Higgs doublets,
achieving the MSSM spectrum. On the other hand, the weak mixing
angle at the unification scale is not $\frac3{8}$. Various mass
scales in addition to the different compactification radii may
enable us to fit the mixing angle to the observed one at the
electroweak scale. A detail analysis of the model for the $R$-parity
problem, weak mixing angle, compactification radii, $D$ and $F$ flat
directions, and Yukawa couplings will be discussed elsewhere.

\acknowledgments{ I thank K.-S. Choi, I. W. Kim and B. Kyae for helpful
discussions. This work is supported in part by the KRF Grants, No.
R14-2003-012-01001-0 and No. KRF-2005-084-C00001.
 }



\end{document}